\documentclass[prl,twocolumn,amsmath,amssymb,showpacs,superscriptaddress]{revtex4-1}
\usepackage{graphicx}
\usepackage{natbib}
\usepackage{bm}
\usepackage{amsfonts}
\usepackage{latexsym}

\usepackage[usenames]{color}
\usepackage{ulem}

\begin{document}
\title{Resonant Control of Interaction Between Different Electronic States}

\author{Shinya Kato}
\email{shinya\_k@scphys.kyoto-u.ac.jp}
\affiliation{Department of Physics, Graduate School of Science, Kyoto University, Kyoto 606-8502, Japan}

\author{Seiji Sugawa}
\affiliation{Department of Physics, Graduate School of Science, Kyoto University, Kyoto 606-8502, Japan}

\author{Kosuke Shibata}
\affiliation{Department of Physics, Graduate School of Science, Kyoto University, Kyoto 606-8502, Japan}

\author{Ryuta Yamamoto}
\affiliation{Department of Physics, Graduate School of Science, Kyoto University, Kyoto 606-8502, Japan}

\author{Yoshiro Takahashi}
\affiliation{Department of Physics, Graduate School of Science, Kyoto University, Kyoto 606-8502, Japan}
\affiliation{JST, CREST, Chiyoda-ku, Tokyo 102-0075, Japan}

\date{\today}

\begin{abstract}
We observe a magnetic Feshbach resonance in a collision between the ground and metastable states of two-electron atoms of ytterbium (Yb).
We measure the on-site interaction of doubly-occupied sites of an atomic Mott insulator state in a three-dimensional optical lattice as a collisional frequency shift in a high-resolution laser spectroscopy.
The observed spectra are well fitted by a simple theoretical formula, in which two particles with an s-wave contact interaction are confined in a harmonic trap.
This analysis reveals a wide variation of the interaction with a resonance behavior around a magnetic field of about 1.1 Gauss for the energetically lowest magnetic sublevel of ${}^{170}$Yb, as well as around 360 mG for the energetically highest magnetic sublevel of ${}^{174}$Yb.
The observed Feshbach resonance can only be induced by an anisotropic interatomic interaction.
This novel scheme will open the door to a variety of study using two-electron atoms with tunable interaction. 
\end{abstract}

\maketitle

A microscopic property of a low energy binary collision determines a macroscopic behavior of an ultracold dilute atomic gas.
Tuning the interaction between the atoms, one of the most fascinating aspect in this system, lies at the heart of recent numerous experimental progresses\,\cite{2008:Bloch:QMPReview}: the formation of ultracold molecules\,\cite{2003:Jin:FermiMol,Greiner2003,Jochim2003,Zwierlein2003}, a Bose--Einstein condensate (BEC) to a Bardeen--Cooper--Schrieffer (BCS) crossover with fermionic gases\,\cite{Regal2004,Zwierlein2004,Zwierlein2005}, Efimov trimer states\,\cite{2006:Grimm:Efimov,Barontini2009,Williams2009}, and so on.
So far, magnetically and optically tuned Feshbach resonances have been utilized\,\cite{2006:Julienne:revPA,2010:Chin:revFR}, in which the interatomic interaction can be resonantly  controlled through the coupling between an open channel and a closed channel.

\begin{figure}[htbp]
   \centering
   \includegraphics[width=8 cm]{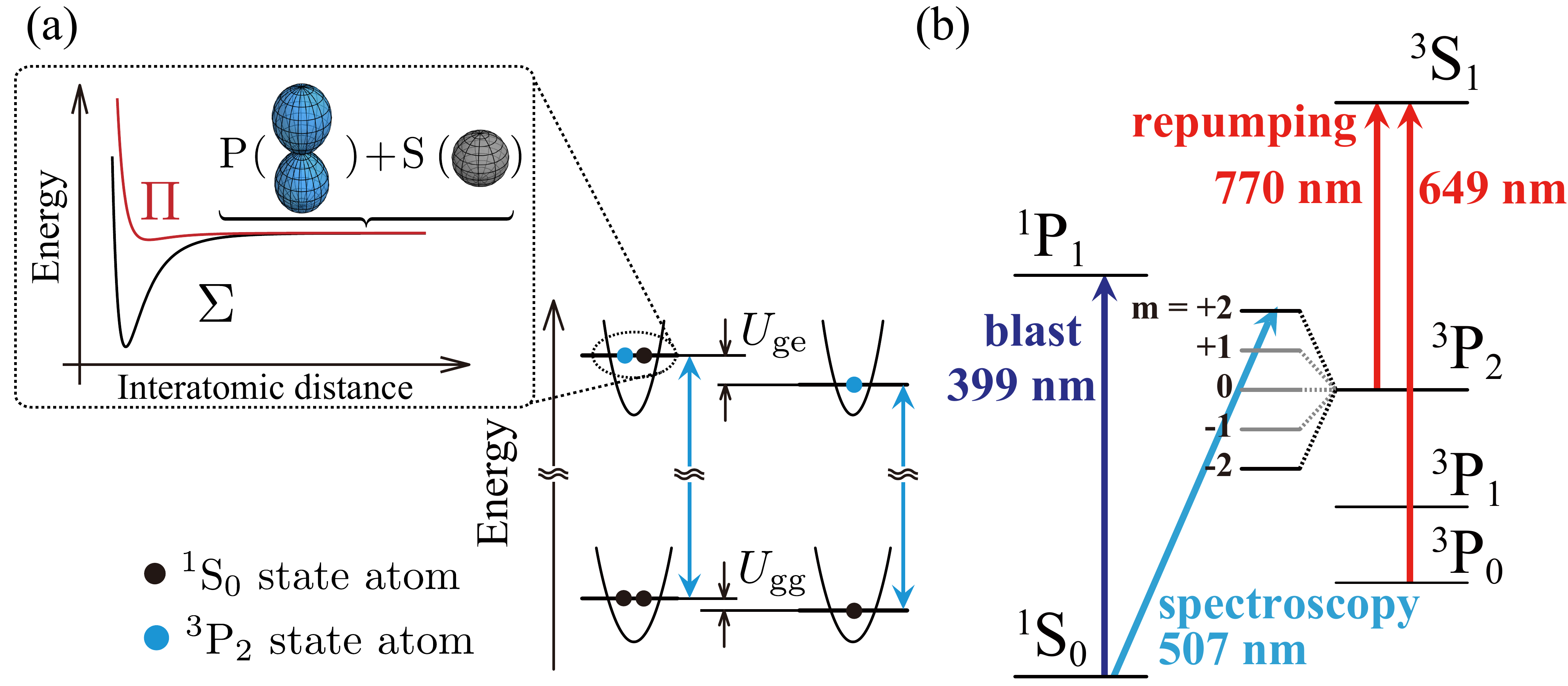}
   \caption{(Color online)
   (a) Energy diagram of singly- and doubly-occupied sites in an optical lattice.
The doubly occupied site has on-site-energy-shifts $U_{gg}$ and $U_{ge}$ due to the interatomic interaction between two ground state atoms and  between  ground and excited state atoms, respectively. 
These energy shifts are revealed by a resonance shift of an excitation spectrum of the doubly-occupied sites compared to that of the singly-occupied sites.
The interatomic potential between the atoms in the ${\rm{}^1S_0}$ and ${\rm{}^3P_2}$ states reflects the anisotropy interaction between them.
(b) Relevant energy levels of an Yb atom for the spectroscopy.
}
   \label{fig:intro}
\end{figure}

Until now all the reported Feshbach resonances, magnetic or optical, were limited to the collision between two electronically ground state atoms.
In this paper, we extend the possibility of the Feshbach resonance to the resonant control of the interaction between electronically excited and ground state atoms.
We successfully observed magnetic Feshbach resonances in collisions between the ground ${\rm {}^1S_0}$ state and the metastable ${\rm{}^3P_2}(m_J=\pm 2)$ states of two-electron atoms of ytterbium (Yb).
Here $m_J$ denotes the magnetic quantum number in the ${\rm{}^3P_2}$ state.
See Fig.\,\ref{fig:intro} for a schematic view of the energy levels of the colliding atoms in the lattice.
The interatomic interaction is directly determined by a high-resolution laser spectroscopy of an atomic Mott insulator state in a three-dimensional optical lattice with the ultranarrow optical ${\rm {}^1S_0\leftrightarrow {}^3P_2}$ transition.
A wide variation of the interaction with a resonance behavior is observed around a magnetic field of about 1.1 Gauss for the energetically-lowest magnetic sublevel $m_J=-2$ of ${}^{170}$Yb, as well as around 360 mG for the energetically highest magnetic sublevel $m_J=+2$ of ${}^{174}$Yb.
The observed resonances are only possible to explain by a recently discussed anisotropic-interatomic-interaction-induced Feshbach resonance mechanism\,\cite{Kotochigova2011,Petrov2012} which can couple different partial waves.
Since the controlled collision between the long-lived metastable ${\rm{}^3P}$ state and the ground ${\rm {}^1S_0}$ state of the alkaline-earth-metal-like atoms has been explored as a useful platform for quantum computing and quantum simulation\,\cite{2008:Daley,2009:Gorshkov,2009:Shibata,Uetake2012}, our results provide a new possibility in such key applications.

Our experiment starts from a preparation of the ultracold ${\rm{}^{174}}$Yb or ${\rm{}^{170}}$Yb BEC, and the detailed procedure is described in Ref.\,\cite{2012:Kato:3P2MRI}.
The produced BEC is loaded into a three-dimensional optical lattice with lineally ramping up the intensity of the lattice laser.
When the lattice depth reaches the final value of 15 ${\rm E_r}$ for ${\rm{}^{174}Yb}$ and 25 ${\rm E_r}$ for ${\rm{}^{170}Yb}$, the systems are in a Mott insulator regime.
Here ${\rm E_r}$ is a recoil energy by a lattice photon.

\begin{figure}[htbp]
   \centering
   \includegraphics[width=8.6 cm]{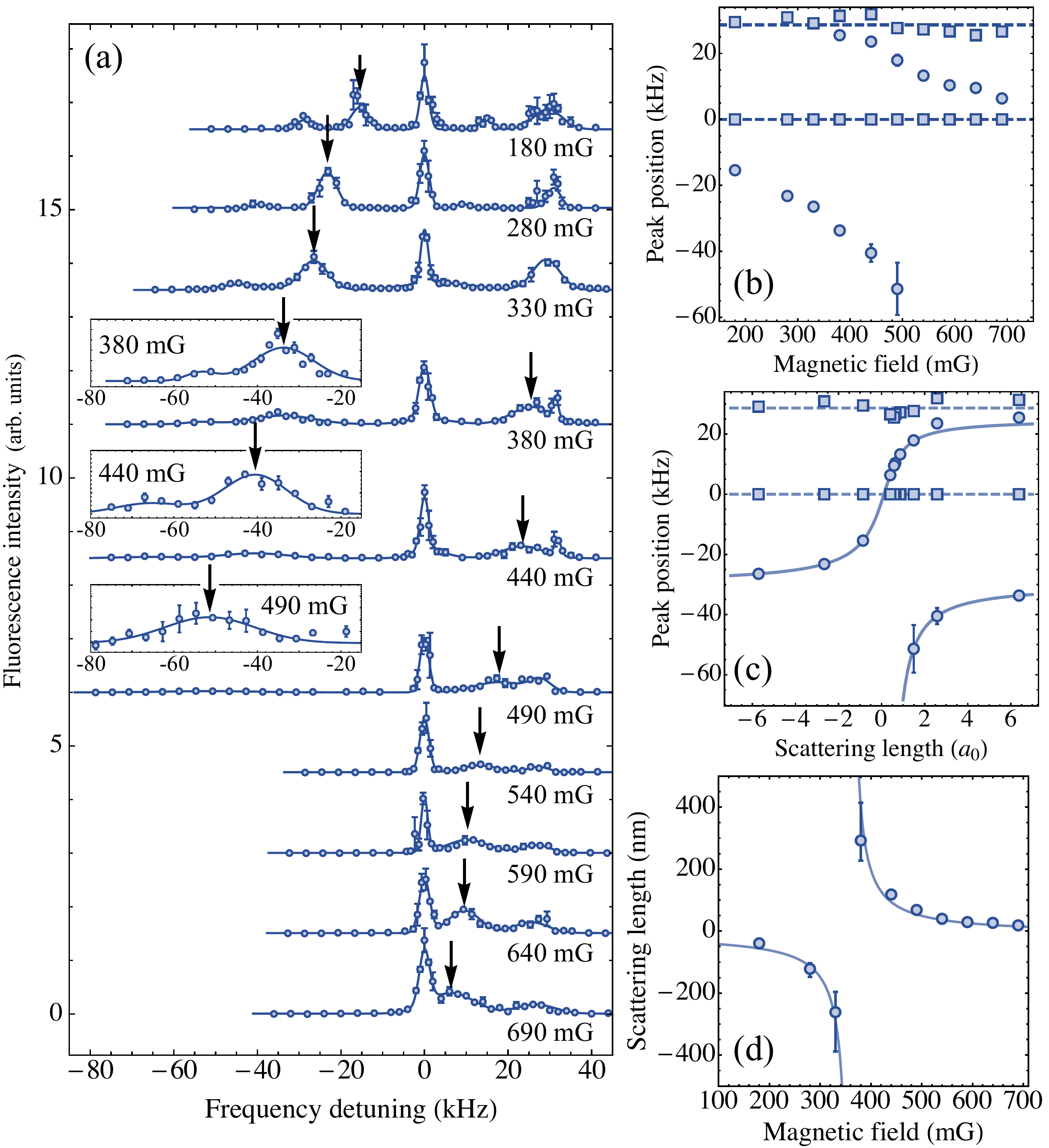} 
   \caption{(Color online)
   (a)Excitation spectra for ${\rm {}^3P_2}(m_J=+2)$ state of ${}^{174}$Yb at various magnetic field strengths below 1 Gauss.
   The insets show enlarged views of low frequency side of the corresponding spectra.
   The arrows indicate the resonances from the doubly-occupied sites.
   The Zeeman shift in each spectrum is subtracted and the spectra are shifted vertically for clarity.
   The solid lines denote the fits with Gaussian functions for each of resonances(see text).
   Plots of resonance peak positions (b) as a function of a magnetic field, and (c) as a function of a scattering length $a_{ge}$ evaluated by using the analytic formula Eq.(\,\ref{eq:cir}).
   The rectangles denote the resonances for the singly occupied sites and excitations to the higher vibrational state, and the circles for the doubly occupied sites.
   The analytical curve based on Eq.(\,\ref{eq:cir}) is indicated by the solid line, and zero and the trap frequency are indicated by the dashed lines.
   The error bars in (b) and (c) represent the standard error of the Gaussian fits.
   (d) Evaluated scattering length $a_{ge}$ as a function of a magnetic field.
   The solid line denotes the function $a_{ge}(B)=a_{bg}-(a_{bg}\Delta)/(B-B_0)$, where $a_{bg}=$ -10 nm is the background scattering length, $\Delta=$830 mG parametrizes the width of the resonance, and $B_0=$ 360 mG is the magnetic field at the resonance center.
}
   \label{fig:main}
\end{figure}

Each site with commensurate fillings realized in the Mott insulator is regarded as an isolated few body system.
We perform a high-resolution laser spectroscopy of the Mott insulator state by using the ultranarrow magnetic quadrupole ${\rm ^1S_0\leftrightarrow ^3P_2}$ transition with the natural linewidth of about 10 {\rm mHz} to study the two-body collisional property of the metastable ${\rm {}^3P_2}$ and ground ${\rm {}^1S_0}$ states.
See Fig. \ref{fig:intro}(b) for relevant energy levels.
The detail of the procedure of the spectroscopy is described in Ref.\,\cite{2012:Kato:3P2MRI}, and here we briefly summarize the procedure.
After the preparation of the Mott insulator, a portion of the ground state atoms is directly excited to the ${\rm{}^3P_2}$ state by 0.25$\sim$1 ms laser pulse whose wavelength is 507 nm.
The atoms remaining in the ground state are blasted out from the trap with  0.2$\sim$0.3 ms laser pulse which is resonant to the electric dipole ${\rm {}^1S_0\leftrightarrow{}^1P_1}$ transition.
The atoms in the excited state are repumped to the ground state via the ${\rm {}^3S_1}$ state by simultaneous applications of two laser pulses which are resonant to the ${\rm {}^3P_2\leftrightarrow {}^3S_1}$ and ${\rm {}^3P_0\leftrightarrow {}^3S_1}$ transitions with the duration of 0.5$\sim$1 ms.
Finally, the repumped atoms are recaptured by a magneto-optical trap (MOT) with the ${\rm {}^1S_0\leftrightarrow {}^1P_1}$ transition for ${}^{174}$Yb or by optical molasses with the ${\rm {}^1S_0\leftrightarrow {}^3P_1}$ transition for ${}^{170}$Yb.
The fluorescence from the MOT or the molasses is detected to measure the number of the repumped atoms.

In Fig.\,\ref{fig:main}(a), we show the excitation spectra obtained with the above method for ${\rm {}^3P_2}(m_J=+2)$ state of ${}^{174}$Yb at various magnetic field strengths below 1 Gauss.
Note that the frequency offset due to the Zeeman shift is already subtracted in Fig.\,\ref{fig:main}(a), and the zero frequency in each spectrum corresponds to the resonance from the singly occupied sites. 
The spectra show additional peaks corresponding to the resonances from the doubly- and triply-occupied sites.
The assignment of these peaks is confirmed by the observations with different total number of atoms, in which the peaks for multiply-occupied sites only appear for large enough number of atoms.
In addition, the excitation peaks to the higher vibrational state in the optical lattice are observed on the higher frequency side.

The interatomic interaction between the ${\rm ^1S_0}$ and ${\rm ^3P_2}$ state manifests itself in the spectrum as a filling-dependent resonance frequency shift\cite{2006:Campbell:MottShell_MIT}, as schematically depicted in Fig.\,\ref{fig:intro}(a).
In the following analysis of evaluating the scattering length, we focus on the resonances from doubly occupied sites which are indicated by arrows in Fig.\,\ref{fig:main}(a). 
In Fig.\,\ref{fig:main}(b), the peak positions are plotted as a function of the magnetic field.
The resonance frequency shift between the singly- and doubly-occupied sites is simply given by the difference of the interatomic interaction between the ground states $U_{gg}$ and that between the ground and excited states $U_{ge}$.
One can clearly see that the frequency separation drastically changes when the magnetic field strength changes.
This behavior directly indicates the change of the interatomic interaction with a magnetic field.
Here the interatomic interaction between the ground states $U_{gg}$ is described as
\begin{equation}
 U_{gg}=\sqrt{\frac{8}{\pi}}k_L a_{gg} E_r s^{3/4}
 \label{eq:Ugg}
\end{equation}
 where $a_{gg}$ is the scattering length, $k_L$ is the wavenumber of the laser for the optical lattice, and $s$ is the potential depth of the lattice divided by $E_r$.
Since $a_{gg}$ does not change with a magnetic field and is precisely determined to be a constant value of 5.55 nm\,\cite{2008:Kitagawa2PAYb-sc}, the observed variation of the resonance peak positions are attributed to the variation of $U_{ge}$.

In order to evaluate the scattering length $a_{ge}$ from $U_{ge}$, we utilize a theoretical formula given in Ref.\,\cite{1998:Busch:CIR} where
 two particles with a contact interaction are confined in an isotropic harmonic trap.
This is important because we cannot use the simple formula like Eq.(\,\ref{eq:Ugg}) when the scattering length becomes close to the harmonic oscillator length $a_{\rm ho}=\sqrt{\hbar/(m\omega)}$.
Here $m$ is the mass of the atom and $\omega$ is the mean trap frequency.  
The interatomic interaction $U_{ge}$ is related with the scattering length $a_{ge}$ by the following analytical formula\,\cite{note1}, 
\begin{equation}
\frac{a_{\rm ho}}{a_{ge}}=\sqrt{2}\frac{\Gamma(-U_{ge}/(2\hbar\omega)+3/4)}{\Gamma(-U_{ge}/(2\hbar\omega)+1/4)},\label{eq:cir}
\end{equation}
where  $\Gamma(x)$ is the Gamma function.

The results of the analysis is shown in Fig.\,\ref{fig:main}(c) where the resonance peak positions are plotted as a function of the evaluated scattering length $a_{ge}$ as well as the energy spectrum calculated from Eq.\,(\ref{eq:cir}).
In the analysis, we use the lower frequency peaks for the magnetic field up to 490 mG, and higher frequency peaks for the magnetic field larger than 490 mG.
Figure \ref{fig:main}(d) shows the magnetic field dependence of the evaluated scattering length $a_{ge}$, which clearly shows a resonance behavior at the magnetic field of about 360 mG.
The scattering length $a_{ge}$ is widely tuned from -260 nm to 300 nm.

\begin{figure}[htbp]
   \centering
   \includegraphics[width=8.6cm]{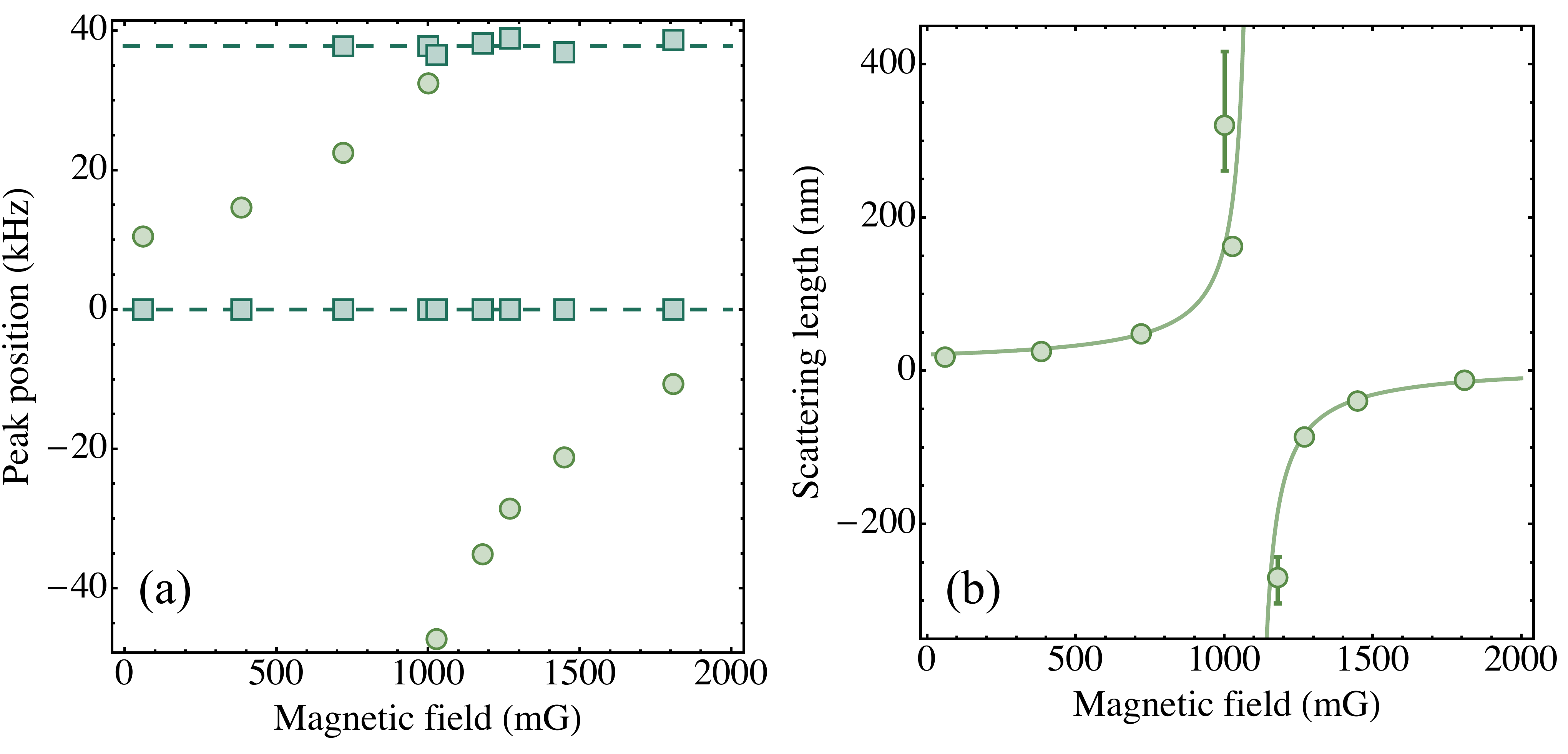} 
   \caption{(Color online)
   (a) Resonance peak positions for ${\rm{}^{170}Yb}$ (${\rm{}^1S_0\leftrightarrow{}^3P_2} (m_J=-2)$) as a function of a magnetic field.
   Circles and rectangles denote the resonances in the same manner in Fig.\,\ref{fig:main}(b).
   (b) Evaluated scattering length $a_{ge}$ between ${\rm{}^{170}Yb}$ atoms in the ${\rm{}^1S_0}$ and ${\rm{}^3P_2}(m_J=-2)$ states.
   The solid line denotes the same form of function $a_{ge}(B)$ as introduced in Fig.\,\ref{fig:main}(d) with $a_{bg}=$ 7 nm, $\Delta=$ 2.2 G, and $B_0=$ 1.1 G.
   }
   \label{fig:170}
\end{figure}

We perform similar measurements for ${\rm {}^3P_2}(m_J=-2)$ state of ${}^{170}$Yb.
The variation of the spectral peaks for the doubly-occupied sites is similarly obtained for magnetic field strengths below 2 Gauss as shown in Fig.\,\ref{fig:170}(a).
From the spectral peaks, we similarly evaluate the scattering length $a_{ge}$ based on Eq. (2).
The result is shown in Fig. \,\ref{fig:170}(b), which again clearly shows the resonance behavior at the magnetic field of about 1.1 G, and $a_{ge}$ is widely tuned from -270 nm to 320 nm.
Note that $a_{gg}$ equals to 3.38 nm for ${}^{170}$Yb\,\cite{2008:Kitagawa2PAYb-sc} .

Here we discuss the mechanism of the observed resonant variation of the scattering length.
First we consider the case of ${}^{170}$Yb.
We note that an isotropic electronic interaction, which is dominant in most cases\,\cite{2010:Chin:revFR}, is not enough to induce a Feshbach resonance in the present case.
This is understood from the conservation of a total angular momentum.
In a collision in the presence of a magnetic field, only the projection $M=m_1+m_2+m_l$ along the magnetic field is conserved\,\cite{2010:Chin:revFR}, where
$m_1$ and $m_2$ are magnetic quantum numbers of two atoms' internal angular momenta $f_1$ and  $f_2$, respectively, and  
 $m_l$ is that of an orbital angular momentum $l$ representing a partial wave of colliding atoms.
The open (entrance) channel ${\rm{}^3P_2}, m_J=-2+{\rm{}^1S_0}$ is an s-wave channel, and therefore it has $M=m_J=-2$.
In the present system of no hyperfine structure, any candidate of closed channels has $m_1+m_2=m_J > -2$, and the difference should be compensated by non-zero $m_l$ of a higher partial wave $l>0$. 
However,
the lack of a spin-spin dipole interaction and the second-order spin-orbit interaction between the ${\rm{}^1S_0}$ and ${\rm{}^3P_2}$ states prohibits the coupling between different partial waves in the case of an isotropic electronic interaction, and therefore, there is no closed channel which can be coupled with the entrance channel.

\begin{figure}[htbp]
   \centering
   \includegraphics[width=8.6cm]{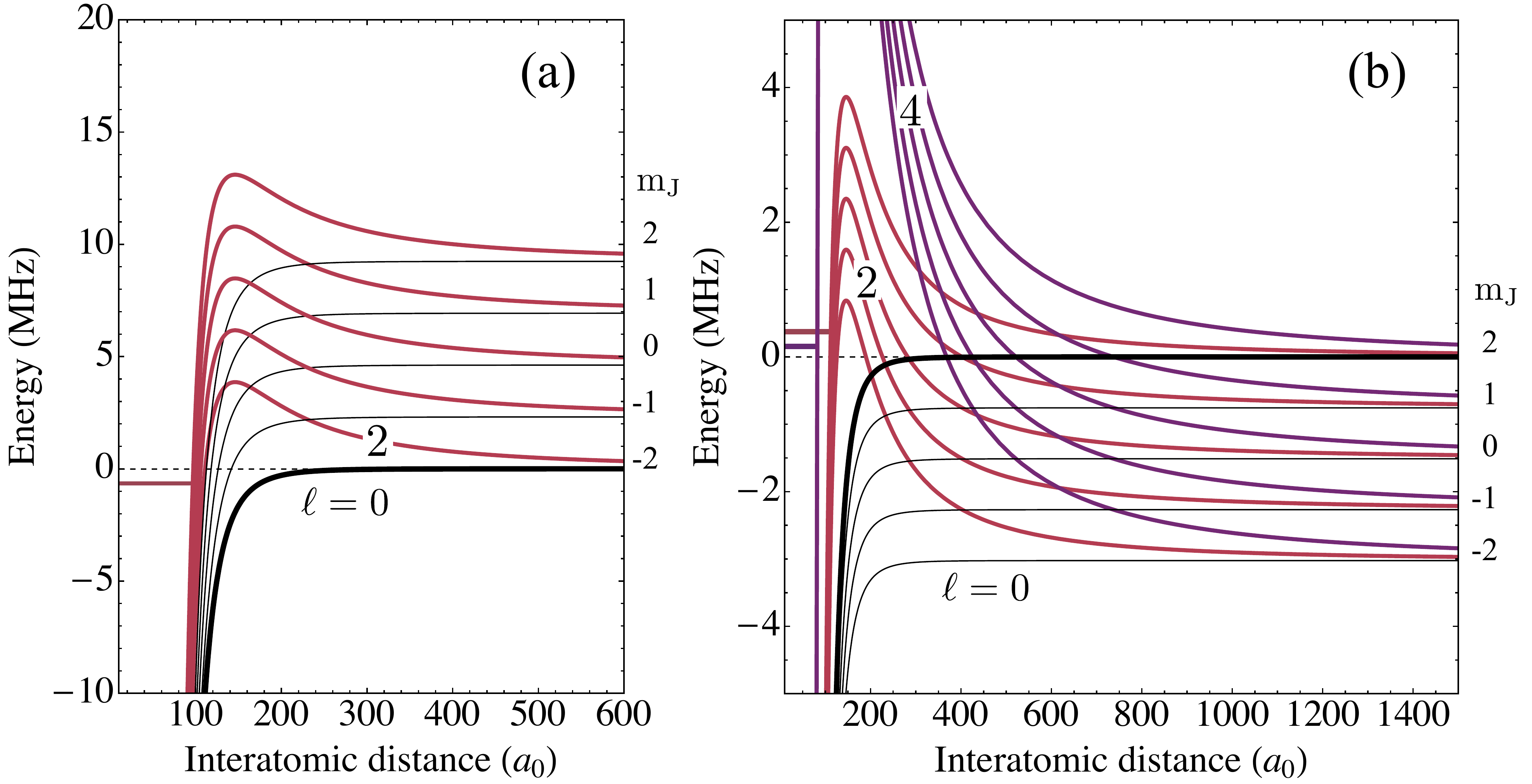} 
   \caption{(Color online)
   Interatomic potential with an anisotropic electronic interaction including the Zeeman shift.
   The magnetic fields are (a) 1.1 G and (b) 360 mG, corresponding to the resonance values of ${}^{170}$Yb and ${}^{174}$Yb.
   Dashed lines with zero energy indicate the energy of the entrance channel for each case.
   In this calculation, we take the values of the Van der Waals coefficients of $C_6^\Sigma$ and $C_6^\Pi$ as 3500 a.u. and 2500 a.u., respectively, as typical ones.
   Here $a_0$ is the Bohr radius.
}
   \label{fig:pot}
\end{figure}

However, for our case of the ${\rm{}^1S_0}+{\rm{}^3P_2}$ collision system, which connects to $\Sigma$ and $\Pi$ ${\rm{}^1S_0}+{\rm{}^3P_2}$ potentials at a short internuclear distance\,\cite{Krems2003}, there is an 
anisotropic interaction which couples the different partial waves with $\Delta l=2$.
The mechanism of the anisotropic-interaction-induced Feshbach resonance is recently discussed for magnetic atoms in Ref.\,\cite{Kotochigova2011,Petrov2012}.
Figure\,\ref{fig:pot}(a) shows the relevant energy levels.
Magnetic sublevels of $m_J > -2$ with the higher partial wave $l=2$ can in general form a bound state inside the centrifugal barrier, as schematically shown by a horizontal solid line in Fig.\,\ref{fig:pot}(a).
Among many states, that with $(m_J, m_l)=(-1,-1)$ or $(0,-2)$ can have a total magnetic quantum number $M=-2$, and thus can be coupled with the open channel, which results in the Feshbach resonance.
Therefore, the observed variation of the scattering length can be consistently explained in terms of the anisotropy-induced Feshbach resonance.
Note that the ${\rm{}^3P_2}(m_J=-2)$ state does not suffer from the Zeeman-sublevel changing collision, and thus it is encouraging to use this Feshbach resonance for controlled collision in a variety of proposed applications in quantum computation and quantum simulation.

In the case of ${}^{174}$Yb, the resonant variation of the scattering length is observed for the ${\rm{}^3P_2}(m_J=+2)$ state which is energetically highest among the Zeeman sublevels of ${\rm{}^3P_2}$.
At first glance, this is surprising because all the Feshbach resonances previously observed have closed (resonance) channels which are energetically higher than the open channel at a long internuclear distance, and such channels are absent in the present case.
We can, however, provide an explanation of this phenomenon by a combination of the anisotropy-induced Feshbach resonance and a shape resonance.
The interplay between the ${\rm{}^1S_0}-{\rm{}^3P_2}$ interatomic interaction along with the centrifugal potential barrier associated with a higher partial wave can form a bound state within the potential barrier above the dissociation threshold, which is known as a shape resonance. 
A possible energy level of the shape resonance is schematically depicted as the horizontal solid line in Fig.\,\ref{fig:pot}(b).
When we consider a shape resonance for lower magnetic sublevels $m_J<+2$, it is not surprising that the energy of such a bound state can be close to or higher than that of the open channel of $m_J=+2$ which is set to zero in Fig.\,\ref{fig:pot}(b). 
Note that the observed magnetic field dependence of the scattering length is consistent with this scenario:
due to the lower magnetic moment of the closed channel compared with the open channel, the energy of the closed channel can be lower than that of the open channel at a high-field side of the resonance, and can be higher at a low-field side.

In conclusion, we observe the magnetically tuned interatomic interaction between different electronic orbitals through the high-resolution laser spectroscopy of the Mott insulator in the 3D optical lattice.
The experimental observations are analyzed by the analytical solution of interacting two atoms in the harmonic trap.
The evaluated scattering length shows a resonant variation around the magnetic field of about 1.1 Gauss for $m_J=-2$ of ${}^{170}$Yb, as well as around 360 mG for $m_J=+2$ of ${}^{174}$Yb.
While the lack of hyperfine structure in the ${\rm{}^3P_2}$ and ${\rm{}^1S_0}$ states and also a spin-spin dipole interaction and the second-order spin-spin interaction between these states prohibits a Feshbach resonance in an isotropic electronic interaction case, Feshbach resonances can be induced in our case by an anisotropic electronic interaction\,\cite{Krems2003} which can couple different partial waves\,\cite{Kotochigova2011,Petrov2012}.
A further theoretical calculation is quite helpful to obtain full quantitative explanation of the observed phenomena, including the position and width of the resonances as well as the identification of the magnetic quantum number corresponding to the closed channel. 

Our work will open the door to a variety of studies using two-electron atoms with tunable interaction, especially to quantum computing and quantum simulation\,\cite{2008:Daley,2009:Gorshkov,2009:Shibata,Uetake2012} in which the controlled collision between the metastable ${\rm{}^3P}$ state and the ground ${\rm {}^1S_0}$ state is a key ingredient. 
The anisotropy-induced Feshbach resonances are also expected for other isotopes.  
Especially interesting case is the fermionic isotopes of ${}^{171}$Yb and ${}^{173}$Yb, where the spin-polarized sample in the ${\rm{}^3P}$ state will not suffer from the inelastic collision\,\cite{2010:Yamaguchi:507,Uetake2012} owing to the Pauli exclusion, and we can possibly study BCS pairing of ground and metastable states. 
In addition, we can also expect similar Feshbach resonances for a mixture system of the ground state of alkali-metal atoms and the metastable state of alkali-earth-metal atoms\,\cite{Hara2011, Hansen2011,Zuchowski2010,Brue2012}.

\begin{acknowledgments}
We acknowledge Y. Yoshikawa for his experimental assistance at the early stage of the study.
We also acknowledge useful discussions with T. Tscherbul, P. Zang, R. Krems, J. Hutson, S. Uetake, and  J. M. Doyle.
This work is supported by the Grant in-Aid for Scientific Research of JSPS (No. 18204035,21102005C01 (Quantum Cybernetics)), GCOE Program "The Next Generation of Physics, Spun from Universality and Emergence" from MEXT of Japan, and World- Leading Innovative R\&D on Science and Technology (FIRST).
\end{acknowledgments}

\begin{thebibliography}{30}
\bibitem [{not()}]{note1}%
  \BibitemOpen
  \href@noop {} {\ }\bibinfo {note} {It is noted that the trap frequencies for
  the ${\rm {}^1S_0}$ state and the ${\rm{}^3P_2}(m_J=\pm2)$ state are
  $2\pi\times 26.5\ {\rm kHz}$ and $2\pi\times 28.5\ {\rm kHz}$, respectively,
  and not the same in general due to the different polarizability for these
  states at the wavelength of the optical trap and optical lattice. Though the
  difference between the trap frequencies causes the mixing of the
  center-of-mass motion and relative motion, however, this mixing effect to the
  eigenenergy is negligible within the first order perturbation. Therefore, we
  use this model without considering the mixing effect and adopt the average of
  these two trap frequencies for the calculation.}\BibitemShut {Stop}%
\end{thebibliography}

%

\end{document}